# An Innovative Brain-Computer Interface Interaction System Based on the Large Language Model

Jing Jin[a,b], Yutao Zhang [a], Ruitian Xu [a], Yixin Chen [a]

[a]Key Laboratory of Smart Manufacturing in Energy Chemical Process, Ministry of Education, East China University of Science and Technology, Shanghai 200237, China
[b]Center of Intelligent Computing, School of Mathematics, East China University of Science and Technology, Shanghai 200237, China

**Abstract**

Recent advancements in large language models (LLMs) provide a more effective pathway for upgrading brain-computer interface (BCI) technology in terms of user interaction. The widespread adoption of BCIs in daily application scenarios is still limited by factors such as their single functionality, restricted paradigm design, weak multilingual support, and low levels of intelligence. In this paper, we propose an innovative BCI system that deeply integrates a steady-state visual evoked potential (SSVEP) speller with an LLM application programming interface (API). It allows natural language input through the SSVEP speller and dynamically calls large models to generate SSVEP paradigms. The command prompt, blinking frequency, and layout position are adjustable to meet the user's control requirements in various scenarios. More than ten languages are compatible with the multilingual support of LLM. A variety of task scenarios, such as home appliance control, robotic arm operation, and unmanned aerial vehicle (UAV) management are provided. The task interfaces of the system can be personalized according to the user's habits, usage scenarios, and equipment characteristics. By combining the SSVEP speller with an LLM, the system solves numerous challenges faced by current BCI systems and makes breakthroughs in functionality, intelligence, and multilingual support. The introduction of LLM not only enhances user experience but also expands the potential applications of BCI technology in real-world environments.

## 1. Introduction

Brain-computer interface (BCI) is a technology that directly realizes information transmission and interaction between the brain and external devices [1]. Through a BCI system, human brain activity can be read, decoded, and converted into commands that can be understood and executed by an external device. The core of BCI technology lies in the conversion of electroencephalographic (EEG) signals (usually obtained through non-invasive or invasive methods) into control signals for human-computer interaction [2], [3].

BCI has been widely used in many fields, with neurorehabilitation [4] and medical assistance [5] being one of the most mature applications. Through BIC technology, paralyzed patients, patients with movement disorders, etc. can control external devices (e.g., prosthetics [6], wheelchairs [7], computers [8], etc.) to perform daily life operations, which greatly improves their quality of life. In recent years, brain-computer interfaces (BCIs) have achieved remarkable results in stroke rehabilitation and motor recovery for patients with spinal cord injuries. In particular, there have been preliminary clinical applications in restoring patients' motor functions through BCI systems [9], [10]. In addition, BCI technology is also widely used in home control [11], drone control [12], robotic arm control [13], and other fields [14]. Through brainwave control, users can realize the control of different devices. In addition, the potential of BCI in all aspects of life is gradually being tapped. However, the popularity of BCI systems faces the following limitations:

First, most of the existing BCI systems are mostly limited to some specific functions, such as simple letter spelling [15], gesture control [16], or lightweight device control [17]. Although SSVEP (Steady State Visual Evoked Potential) technology[18] performs well in spelling systems, existing systems appear to be under-functional for complex device control needs, such as smart home, drone manipulation, or complex task planning. Second, most of the existing SSVEP paradigm designs rely on pre-set flicker frequencies and position layouts, making it difficult to dynamically generate personalized control interfaces [19]. This makes users need to rely on fixed and limited interfaces to interact in different scenarios, reducing the flexibility and practicality of the system. Third, most of the existing BCI systems are limited to specific languages [20], especially non-English speaking users who often lack complete localization support when using them. Cross-language support is very limited, especially with the lack of integration with natural language understanding and generation. Last, although some BCI systems can perform relatively simple commands, such as controlling electrical switches or robot movement, the systems are unable to reason based on complex semantics, automatically generate interfaces adapted to different tasks, and lack deep integration with intelligent assistants or large language models. These are all issues that still need to be addressed in the current BCI system.

The rise of LLM represents a breakthrough in artificial intelligence, showcasing powerful natural language processing capabilities [21]. By leveraging vast amounts of textual data for deep learning, LLMs can generate language output that aligns with contextual logic, demonstrating significant potential for human-like interaction across a wide range of application scenarios [22]. Advanced LLMs, such as GPT-4 [23], PaLM [24], LLaVA [25], and Wenxin Yiyan, have continued to improve their understanding of complex contexts and expand their support for multiple languages. This enables LLM to adapt to more complex task-processing requirements, showing great potential in cross-domain applications. By integrating LLMs, BCI systems can automatically analyze and perform complex tasks, overcoming the limitations of traditional brain-computer interfaces and providing a more natural and convenient interactive experience.

Based on the above challenges and inspirations, this paper introduces a novel brain-computer interface system that solves the shortcomings of existing systems in terms of flexibility, intelligence, and multilingual support by combining the SSVEP speller with LLM. The contributions of this paper include:

- The system enters commands through the SSVEP speller and utilizes them to invoke the LLM API. LLM possesses powerful natural language understanding and generation capabilities and can generate control interfaces adapted to different scenarios input by the user.
- The SSVEP interface of the system is no longer fixed, but a new paradigm generated by the LLM based on the number of devices, device names, and other factors. The blinking color block text, blinking frequency, and layout position of this paradigm are dynamically generated according to specific scenes, which improves the flexibility and adaptability of the system.
- The system is compatible with more than ten languages through LLM's multi-language support. Users can input commands in any of the supported languages, and the big model will generate the corresponding control interface according to the input language. This greatly expands the scope of application of the system, especially in internationalized application scenarios, providing users with a seamless cross-lingual experience.
- The system is not limited to home appliance control, but is also capable of controlling devices such as robotic arms and drones, providing compatibility with a wide range of application scenarios. With the semantic understanding capability of the large model, users can input more complex task instructions, and the system can automatically generate the corresponding SSVEP control interface based on the instructions, which is suitable for a wide range of task execution.

The rest of this paper is organized as follows: in Section 2, we describe this work involving data acquisition and processing, dynamic SSVEP page generation, multilingual support, and cross-domain applications. A demonstration of the working pages of the system is carried out in Section 3. Finally, the research work is summarized and an outlook for future research is presented in Section 4.

## 2. Materials and Methods

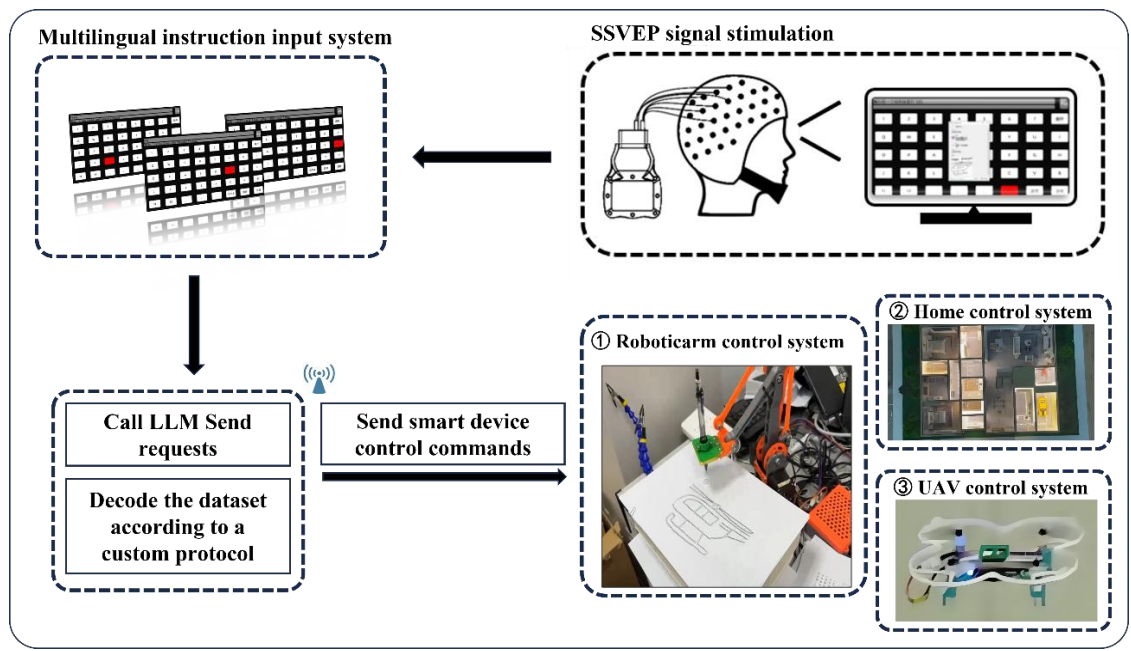

Fig. 1. System implementation flowchart

The proposed system integrates the SSVEP speller with LLM API to enable multidimensional control of intelligent devices. Fig. 1 shows the overall workflow of the system. To provide a comprehensive understanding of the concrete implementation of this system, this section explains each technical detail in depth, covering aspects such as data acquisition, SSVEP signal processing, multilingual support, dynamic SSVEP interface generation, and cross-domain applications.

*2.1. Data acquisition and signal processing*

Data acquisition is a critical step in achieving brain-computer interaction, as it involves extracting electrical signals from the user's brain that are related to brain activity. In this system, the SSVEP speller is used to implement brain-computer interface input, decoding the user's intent based on the brain's response to visual stimuli at different frequencies. Specifically, a wireless 64-channel 1000Hz high sampling rate EEG device is used for data acquisition. This high sampling rate ensures that the system can accurately capture the subtle brainwave responses generated when the user gazes the flashing block at different frequencies, thus avoiding the loss of critical brainwave information due to the low sampling rate.

After the data acquisition is completed, the system preprocesses the EEG signals. The preprocessing step involves removing 50Hz power line noise, as well as eliminating eye movement and muscle artifacts. The system adopts a filter and artifact removal algorithm to ensure the purity and accuracy of the signal and prevent non-target signals from interfering with the analysis results. The system uses Task-Discriminant Component Analysis (TDCA) [26] to decode the collected EEG signals. TDCA is a dimensionality reduction algorithm designed for multi-task learning. TDCA seeks an optimal projection matrix that maps the original high-dimensional data into a lower-dimensional space. In this new space, the similarity of samples within the same task is maximized, while the differences between tasks are amplified. This enhances task discrimination, allowing for precise identification of the visual stimulus frequency the user is focusing on, and generating effective voice commands to invoke the large language model API.

TDCA uses a discriminative model, which is different from traditional strategies that optimize spatial filters for each class individually. Instead, it performs discriminative analysis to search for spatial filters that are relevant to all classes. First, the enhanced training and testing data are formed by combining the original signals with their time-delayed versions. During the individual calibration process, for a given $i$-th trial, $X^{(i)} \in R^{N_{ch} \times N_p}$, $i = 1, 2, \cdots, N_t$, where $N_t$ represents the number of trials and $N_p$ represents the number of sampling points. The enhanced data is then:

$$\tilde{X} = \left[ X^T, X_1^T, \cdots X_l^T \right]^T \tag{1}$$

where $\tilde{X} \in R^{(l+1)N_{ch} \times N_p}$ represents the augmented EEG signal after data enhancement, and $X_l \in R^{N_{ch} \times N_p}$ represents the trial data delayed by $l$ points, i.e., the delayed version of the data from time (data point) $l+1$ to time $N_p + l$. During the augmentation of the test data, to ensure that no data points greater than $N_p$ are included in the test data, any points beyond $N_p$ are filled with zeros.

$$X_l = \left[ X_l', O^{N_{ch} \times l} \right] \tag{2}$$

where $X_l' \in R^{N_{ch} \times (N_p - l)}$ represents the data from time $l+1$ to time $N_p$. The enhanced data is then projected onto the subspace spanned by the reference signal.

$$\tilde{X}_p = \tilde{X} P_i \tag{3}$$

where $P$ is the orthogonal projection matrix for the $i$-th class:

$$P_i = QQ^T \tag{4}$$

where $Q$ comes from the QR decomposition of the sine and cosine reference signal $Y$, corresponding to the $i$-th stimulus frequency $f_i$:

$$Y_i = QR \tag{5}$$

$$Y_i = \begin{bmatrix} sin(2\pi f_i t^T) \\ cos(2\pi f_i t^T) \\ \vdots \\ sin(2\pi N_h f_i t^T) \\ cos(2\pi N_h f_i t^T) \end{bmatrix}^T, t = \left[ \frac{1}{f_s}, \cdots, \frac{N_p}{f_s} \right]^T \tag{6}$$

where $N_h$ is the number of harmonics, and $f_s$ is the sampling rate. Then, for the training and testing data, further secondary data augmentation is performed.

$$X_a = \left[ \tilde{X}, \tilde{X}_p \right] \tag{7}$$

Then, two-dimensional linear discriminant analysis (LDA) is performed on the training trials to find the projection direction that discriminates between all classes. The between-class scatter matrix and within-class scatter matrix are defined as follows:

$$H_b = \frac{1}{\sqrt{N_c}} \left[ \bar{X}_a^1 - \bar{X}_a^a, \cdots, \bar{X}_a^{N_c} - \bar{X}_a^a \right] \tag{8}$$

$$H_w = \frac{1}{\sqrt{N_t}} \left[ X_a^{(1)} - \bar{X}_a^{(1)}, \cdots, X_a^{(N_t)} - \bar{X}_a^{(N_t)} \right] \tag{9}$$

where $\bar{X}^i$ and $\bar{X}^{(i)}$ represent the two-dimensional class centers for the $i$-th class and the $i$-th sample, respectively. The superscript $a$ indicates all classes, and $\bar{X}_a^a$ is obtained by the following formula (10):

$$\bar{X}_a^a = \frac{1}{N_t} \sum_{i=1}^{N_t} X_a^{(i)} \tag{10}$$

Then, the projection direction is derived using the Fisher criterion:

$$\underset{W}{\text{maximize}} \frac{tr(W^T S_b W)}{tr(W^T S_w W)} \tag{11}$$

where the scatter matrices $S_b$ and $S_w$ are represented as:

$$S_b = H_b H_b^T \tag{12}$$

$$S_w = H_w H_w^T \tag{13}$$

Finally, by using the idempotency of the projection matrix, i.e., $P^2 = P$, equation (11) can be rewritten as:

$$\underset{W}{\text{maximize}} \frac{tr\left[W^T H_b' (P_b + I) H_b'^T W\right]}{tr\left[W^T H_w' (P_w + I) H_w'^T W\right]} \tag{14}$$

where $H_b'$ and $H_w'$ are composed of $\tilde{X}$ from equations (8) and (9). For the projection matrices, $P_b = \oplus_{i=1}^{N_c} P_i$, $P_w = \oplus_{i=1}^{N_c} \oplus_{j=1}^{N_b} P_i$, where $\oplus$ denotes the direct sum, and $N_b$ represents the number of blocks.

Finally, based on the obtained general filter $W$, the correlation coefficient between $X_a^{(i)}$ generated from a single trial an $\bar{X}_a^i$ is solved as:

$$\rho_n = \text{corr}\left(W^T X_a^{(i)}, W^T \bar{X}_a^i\right) \tag{15}$$

The target frequency $f_{target}$ is identified based on the stimulus frequency corresponding to the maximum correlation coefficient:

$$f_{target} = \underset{n}{\arg\max} \, \rho_n, \quad i = 1, 2, \cdots, N_f \tag{16}$$

These steps ensure that the SSVEP signals can be quickly and accurately parsed into results that reflect the user's intent, thus driving the system to perform subsequent operations.

*2.2. Multilingual support*

Through the integration with the LLM, the system is compatible with command input in more than ten languages. Users can enter control commands in Chinese, English, French, Spanish, and other languages, with the system automatically detecting and processing the input language. The LLM then generates the appropriate control interface based on the detected language, without additional translation or language conversion. Designing a friendly multilingual interface not only expands the user base, but also provides a convenient and efficient multilingual interactive experience for users with different language backgrounds, and enhances the global adaptability and competitiveness of the system.

*2.3. Dynamic SSVEP interface generation*

Users input commands through the SSVEP speller, which the system sends to the LLM API via an HTTP request. LLM utilizes its advanced natural language processing capabilities to perform semantic understanding of the input, identifying the user's intent. It generates feedback tailored to the current task context based on semantic reasoning, ensuring that the list of devices provided accurately matches the needs of the users. When a user requests home appliance control, the LLM responds with a list of controllable devices (lights, air conditioning, and television).

The traditional BCI system often has a static interface, which can not adapt to the real-time needs of users, but this system uses the device information generated by LLM to create a dynamic control interface, generate device control text, and adjust the flashing frequency, color and layout of each square. It automatically calculates the best layout, positioning each flashing square within the user's field of view for easier selection by staring. The flicker frequency is designed according to the complexity of the device control and user preferences, ensuring that there is enough frequency difference to minimize selection errors. The font size of each square can also be dynamically adjusted to maintain clarity under different lighting conditions and enhance the overall user experience. In addition, LLM provides detailed control tasks for each device, such as turning on or off, adjusting temperature or changing brightness.

*2.4. Cross-domain applications*

The system provides broad cross-domain application compatibility. In addition to home appliance control, it also supports the management of complex devices, such as robotic arms and drones. This versatility is achieved through a common control interface that allows the system to interact with a variety of devices and generate customized SSVEP interfaces based on the specific control requirements of each device. The automated reasoning and dynamic generation capabilities of LLM make the system suitable not only for simple tasks, but also for more demanding and complex operating environments. Below is a more detailed introduction to its application across various scenarios.

*2.4.1 Home control system application*

The system leverages Home Assistant for automatic device discovery within the local area network (LAN). Home Assistant is an open-source smart home control platform that integrates various protocols (Zigbee, Z-Wave, Wi-Fi) and supports a wide range of smart devices from multiple brands and models. The API of the Home Assistant scans the LAN for connected devices upon startup, collecting information about the type, status, name, IP address, control protocol, and functional descriptions, and so on of all controllable devices. A standardized interface for device discovery will be provided by Home Assistant's API, enabling the system to obtain real-time details. Once device discovery is complete, the system sends the device information to the LLM for parsing and generates a new control interface based on the SSVEP interface's requirements. The JSON data returned by the LLM typically contains the following key elements:

(1) Device name: the name of each controllable device, such as "light", "air conditioner", "TV".
(2) Device functions: the executable operations corresponding to each device, such as "on/off", "adjust temperature", "adjust brightness".
(3) Device status: the current status of each device, such as "on" or "off", which is used to dynamically update the interface.

The system can automatically generate the control interface suitable for the current LAN environment without manual configuration, and users can effectively control the home equipment through SSVEP. By using the automatic generation and reasoning ability of the LLM, the system plays the role of sorting, classification and reasoning of home equipment information, and provides more versatility and usability for the realization of smart home.

*2.4.2 Robotic arm control applications*

When controlling complex robotic arm tasks, the user enters commands through the SSVEP spelling interface, such as "I want to operate the robotic arm" or more specific task instructions, such as "grab the object and place it in the designated position." Once the user submits these natural language commands, the system sends them to the LLM via an API. The LLM processes the input to extract the operation object (the robotic arm) and the task objective (grabbing and placing the object). Not only the basic operation commands, but also the more detailed operation steps need to be inferred based on the context and the current scenario. For example, for a complex task like "grab the object," the model generates sub-tasks such as "moving the robotic arm over the object", "adjusting the gripper's posture", "grabbing the object", and "executing the placement operation". After receiving the instructions from the LLM, the system breaks them down into specific control actions, which are then executed step by step through the robotic arm control module. For instance, the JSON response from the LLM may include the following components:

(1) Movement command: such as "move 5 cm in the positive X-axis direction".
(2) Posture adjustment: such as "rotate the gripper by 45 degrees".
(3) Grabbing instruction: such as "execute grab".
(4) Placing instruction: such as "move to the target position and place".

These control commands are dynamically represented as blinking blocks in the SSVEP interface. The user can select a specific robot arm action by focusing on the corresponding block. For a simple task like "move", the system shows basic orientation control options, while for a more complex task like "grab and place", the system shows additional subtasks and action steps that help the user complete the task in stages.

*2.4.3 UAV control applications*

UAV control often involves complex operations such as flight path planning, speed adjustment, and mission execution (for example, taking photos or transporting items). Users can enter flight commands through the SSVEP spelling interface, such as "fly the drone to a specified location and take a photo" or "plan a flight path". LLM performs semantic analysis on these commands, understands the user's task requirements, and generates corresponding control steps. For example, when the user inputs "fly the UAV to a specified location and take a photo", LLM generates the following control steps based on the environment and task requirements:

(1) Flight path planning: such as "fly 100 meters north".
(2) Speed adjustment: such as "set the flight speed to 5 meters per second".
(3) Task execution: such as "take a photo upon reaching the specified location".

These control steps are displayed through the SSVEP interface, where users can select the flight path, adjust speed, or perform the photo-taking task by focusing on the relevant control blocks. The system ensures the accuracy and intelligence of the UAV task by using the control commands generated by the LLM. For instance, when the user's input is incomplete or unclear, LLM will automatically fills in task details through context reasoning, such as smartly selecting the best flight path or adjusting flight parameters based on environmental conditions during task execution.

## 3. Experimental Results and Analysis

### 3.1. Multitasking Application Scenario Building

Most of the existing brain-computer interface systems are mostly limited to some specific functions such as simple letter spelling, gesture control, or lightweight device control. Although SSVEP (Steady State Visual Evoked Potential) technology performs well in spelling systems, the existing system functions appear to be insufficient for complex device control needs, such as smart home, drone manipulation, or complex task planning. Therefore, after our design, our system can realize the construction of multi-tasking daily application scenarios and dynamically generate personalized control interfaces. The SSVEP interface of this system is no longer fixed, but a new paradigm is generated by the LLM based on the number of devices, device names, and other factors. The blinking color block text, blinking frequency, and layout position of this paradigm are dynamically generated according to the specific scene, which improves the flexibility and adaptability of the system. Users can generate customized interfaces for different scenarios, such as home appliance control, robotic arm operation, or UAV mission planning, to provide a more intuitive interaction experience.

The system's management of complex devices such as appliance control, robotic arm control, and drone manipulation is achieved through a common control interface. The system is capable of interacting with a wide range of devices and generating customized SSVEP interfaces according to the control requirements of different devices. The specific system is presented in the Fig. 2 below:

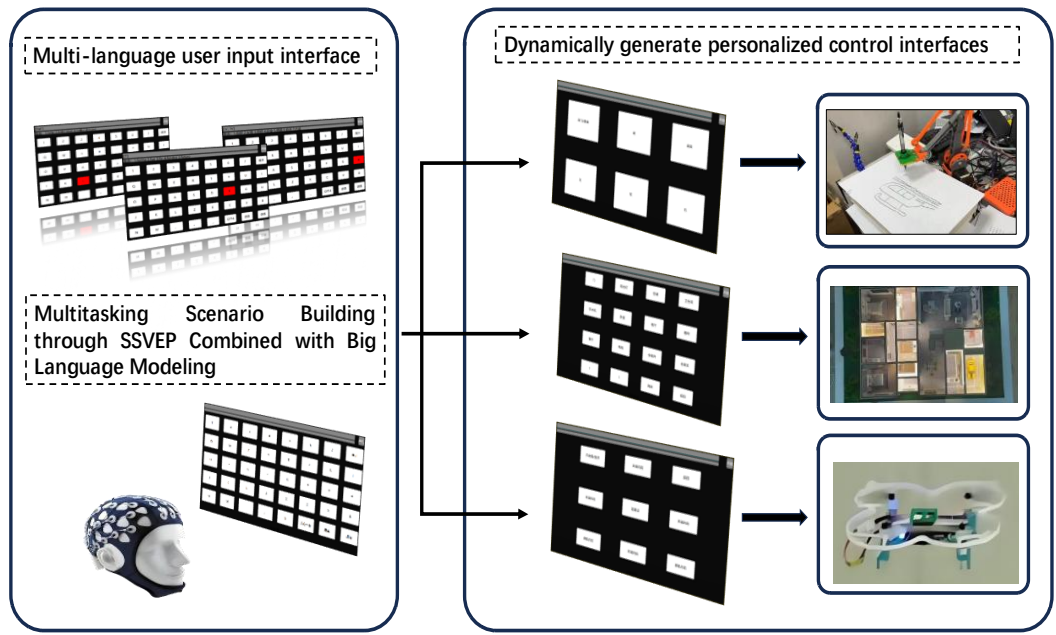

Fig. 2. Multitasking Application Scenario Detail Diagram

When controlling the robotic arm, LLM generates a series of operation options, such as moving, grasping, rotating, and so on. After the user selects the appropriate operation through the SSVEP speller, the system will pass the operation command to the robotic arm control module to execute the specific action. For UAV control, the system generates control options such as flight path selection, speed adjustment, etc., and the user similarly selects the flight mode and plans the mission by gazing at the blinking square.

Unlike traditional control methods, this brain-computer interface-based control does not require manual operation, which greatly improves the efficiency of the user's task execution in complex scenarios.

*3.2. Support for multilingual users*

Existing brain-computer interface systems are mostly limited to specific languages, especially non-English speaking users who often lack complete localization support when using them. Cross-language support is very limited, and especially integration with natural language understanding and generation is lacking. Therefore, after our design, our system can provide support for multilingual user usage. The system can be compatible with more than ten languages through a large model of multilingual support. Users can input commands in any of the supported languages, and the big model will generate the corresponding control interface according to the input language. This greatly expands the scope of the system, especially in internationalized application scenarios, providing users with a seamless cross-lingual experience.

In the project of building a module to support multilingual input, we completed several key aspects through a systematic approach, from data preprocessing to model training and optimization, and then to the implementation of practical applications. The project includes the collection and processing of linguistic data, the selection and training of models, and the implementation of multilingual support extensions.

During the data preprocessing phase, we collected a corpus of texts in multiple languages to ensure that all target languages needed to be supported in the project were covered. We cleaned and normalized the text data using built-in tools, including steps such as removing punctuation marks, converting to lowercase letters, and removing stop words. The purpose of this step is to ensure the purity and consistency of the text data and lay the foundation for the subsequent model training.

During the model training phase, we choose the method based on statistical language modeling and train the processed text data through the n-gram modeling module. We trained multiple n-gram models for different languages, such as 2-gram and 3-gram models, to capture lexical and phrase structural features of different languages. We used the Maximum Likelihood Estimation (MLE) method to calculate the probability distribution of each n-gram to construct the language models.

To enhance the generalization ability of the model, we introduced smoothing techniques, such as Laplace smoothing and Good-Turing smoothing, which deal with low-frequency words and unseen words. These techniques reduce the model bias due to insufficient training data by adjusting the probability distribution and enhancing the performance of the model in real applications.

For multi-language support, we have designed a unified framework that can flexibly add or remove support for different languages. The framework manages models for different languages through language identifiers, and automatically detects the language entered by the user during the input process, to call the corresponding language model for associative input. This design not only improves the scalability and maintainability of the system but also ensures the consistency of input experience in a multilingual environment. Examples in Chinese, Spanish, and English are shown in the Fig. 3.

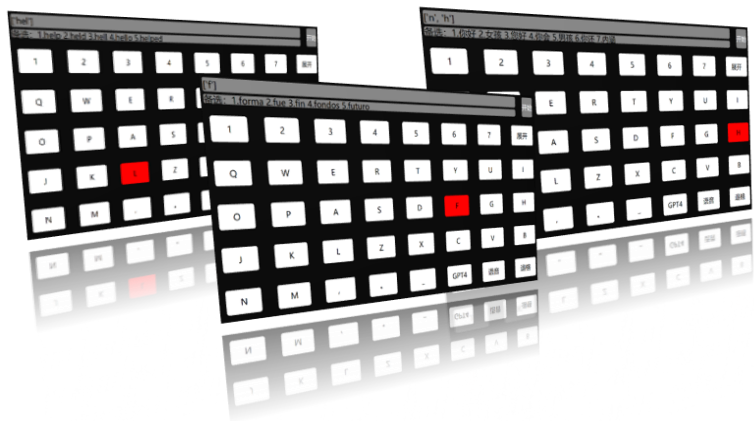

Fig. 3. Multi-language user interface

We have successfully built an associative input module that supports multiple languages and has been put into use in real projects. The module performs well in several application scenarios, which not only improves users' input efficiency but also provides a good cross-lingual user experience. This achievement not only demonstrates the power of the NLTK framework in natural language processing but also sets a new benchmark for associative input technology with multilingual support.

*3.3. Complex semantic reasoning systems*

Although some brain-computer interface systems are capable of executing simpler commands, such as controlling electrical switches or robot movement, the systems are unable to reason based on complex semantics, automatically generate interfaces adapted to different tasks, and lack deep integration with intelligent assistants or large language models. Therefore, after our design, our system can reason about complex semantics and automatically generate commands for different tasks.

We achieve this by designing a set of customized protocols to convert natural language commands into specific operating instructions for robotic arms, drones, and smart home devices. The protocol is designed to be concise and effective in parsing and executing the natural language commands entered by the user, ensuring the reliability and efficiency of the system.

Our system inputs Chinese phrases through the SSVEP speller and utilizes voice commands to call the LLM, which has powerful natural language understanding and generation capabilities and can generate control commands adapted to different scenarios based on the user input phrases.

In the specific implementation process, we first parse the user-input commands using natural language processing techniques. The LLM is responsible for understanding the user's intention and converting it into corresponding device control commands. The parsing steps include recognizing the subject of the command (e.g., robotic arm, drone, smart home device, etc.) and the specific operation (e.g., moving, switching, setting parameters, etc.).

Once the user's instruction intent is recognized, the system generates the corresponding control code. For example, when the user inputs "move robot arm to coordinate (10, 20, 30)", the system will convert it to the instruction format of "$Robot arm(10, 20, 30)". Similarly, commands for drones and smart home devices are converted into a predefined format to ensure that each device can accurately receive and execute commands. To better demonstrate the conversion process of the commands, the following table lists the control commands of several typical devices and their corresponding natural language examples.

Table 1. Typical control commands and their natural language counterparts

| Equipment type | Natural language command | Converted control instructions |
| --- | --- | --- |
| Robot arm | Move the robot arm to coordinates (10, 20, 30) | $Robot arm (10, 20, 30) |
| And another entry | The drone is moving one meter to the left. | $Quadcopter (0, 0, 1, 0) |
| Intelligent Light | Turn on the living room light. | $Lamp (living room, 1) |
| Intelligent Thermostat | Set the temperature to 22 degrees | $Thermostat (set, 22) |
| Intelligent Curtains | Open the bedroom curtains. | $Curtain (bedroom, open) |

The system utilizes custom protocols to achieve efficient conversion of natural language commands to specific operations and ensures system scalability and compatibility through standardized interface design. This achievement not only demonstrates the application potential of our intelligent system in smart device control but also lays a solid foundation for further development in the field of smart home and automation control in the future.

**4. Conclusions and Future Work**

The design of this system is not only an upgrade to existing brain-machine interface (BCI) technology but also a high-level exploration aimed at pushing BCI technology into more extensive, intelligent, and personalized application domains.

(1) Cross-domain intelligent control center: By integrating with LLM, the system is expected to become a cross-domain intelligent control center, playing a significant role in areas such as smart homes, industrial automation, and personal assistant devices. The system's dynamic SSVEP interface generation and intelligent task reasoning provide users with a convenient and efficient interaction method.

(2) Inclusive technology application: With enhanced multilingual support, this system can meet the needs of global users, reducing language barriers and helping more people improve their quality of life through brain-machine interface technology. Additionally, the system's scalability enables it to quickly adapt to new devices and emerging technologies, presenting strong prospects for future applications.

(3) Personalized interactive experience: In the future, based on user's data and preferences, this system can further optimize the generated SSVEP paradigms, achieving truly personalized interaction. Whether for home control, robotic arm task execution, or UAV flight, the system can offer interfaces best suited to users' habits.

(4) Promoting the popularization and intelligent development of BCI technology: Through the innovative design of this system, brain-machine interface technology moves beyond simple interactive operations and advances toward more complex task execution and intelligent reasoning. This system not only enhances the efficiency of existing BCI technology but also lays a foundation for future intelligent applications, making significant contributions to the popularization and intelligent development of BCI technology.

By deeply integrating the SSVEP spell interface and large models, the system addresses many of the pain points of current BCI systems and has achieved breakthroughs in functionality, intelligence, and multilingual support.

**References**

[1] L. F. Nicolas-Alonso and J. Gomez-Gil, "Brain computer interfaces, a review," *sensors,* vol. 12, no. 2, pp. 1211-1279, 2012.

[2] F. Cincotti *et al.*, "Non-invasive brain–computer interface system: towards its application as assistive technology," *Brain research bulletin,* vol. 75, no. 6, pp. 796-803, 2008.

[3] T. Lal *et al.*, "Methods towards invasive human brain computer interfaces," *Advances in neural information processing systems,* vol. 17, 2004.

[4] N. Birbaumer and P. Sauseng, "Brain–computer interface in neurorehabilitation," in *Brain-Computer Interfaces: Revolutionizing Human-Computer Interaction*: Springer, 2010, pp. 155-169.

[5] J. J. Shih, D. J. Krusienski, and J. R. Wolpaw, "Brain-computer interfaces in medicine," in *Mayo clinic proceedings*, 2012, vol. 87, no. 3: Elsevier, pp. 268-279.

[6] D. J. McFarland and J. R. Wolpaw, "Brain-computer interface operation of robotic and prosthetic devices," *Computer,* vol. 41, no. 10, pp. 52-56, 2008.

[7] F. Galán *et al.*, "A brain-actuated wheelchair: asynchronous and non-invasive brain–computer interfaces for continuous control of robots," *Clinical neurophysiology,* vol. 119, no. 9, pp. 2159-2169, 2008.

[8] D. Tan and A. Nijholt, *Brain-computer interfaces and human-computer interaction*. Springer, 2010.

[9] N. Birbaumer, "Breaking the silence: brain–computer interfaces (BCI) for communication and motor control," *Psychophysiology,* vol. 43, no. 6, pp. 517-532, 2006.

[10] M. A. Cervera *et al.*, "Brain-computer interfaces for post-stroke motor rehabilitation: a meta-analysis," *Annals of clinical and translational neurology,* vol. 5, no. 5, pp. 651-663, 2018.

[11] F. Miralles *et al.*, "Brain computer interface on track to home," *The Scientific World Journal,* vol. 2015, no. 1, p. 623896, 2015.

[12] J.-H. Jeong, D.-H. Lee, H.-J. Ahn, and S.-W. Lee, "Towards brain-computer interfaces for drone swarm control," in *2020 8th International Winter Conference on Brain-Computer Interface (BCI)*, 2020: IEEE, pp. 1-4.

[13] M. Y. Latif *et al.*, "Brain computer interface based robotic arm control," in *2017 International Smart Cities Conference (ISC2)*, 2017: IEEE, pp. 1-5.

[14] D. Plass-Oude Bos *et al.*, "Brain-computer interfacing and games," *Brain-computer interfaces: applying our minds to human-computer interaction,* pp. 149-178, 2010.

[15] X. Chen, Y. Wang, M. Nakanishi, X. Gao, T.-P. Jung, and S. Gao, "High-speed spelling with a noninvasive brain–computer interface," *Proceedings of the national academy of sciences,* vol. 112, no. 44, pp. E6058-E6067, 2015.

[16] N. Korovesis, D. Kandris, G. Koulouras, and A. Alexandridis, "Robot motion control via an EEG-based brain–computer interface by using neural networks and alpha brainwaves," *Electronics,* vol. 8, no. 12, p. 1387, 2019.

[17] R. A. Ramadan and A. V. Vasilakos, "Brain computer interface: control signals review," *Neurocomputing,* vol. 223, pp. 26-44, 2017.

[18] Y. Wang, X. Chen, X. Gao, and S. Gao, "A benchmark dataset for SSVEP-based brain–computer interfaces," *IEEE Transactions on Neural Systems and Rehabilitation Engineering,* vol. 25, no. 10, pp. 1746-1752, 2016.

[19] H. Bakardjian, T. Tanaka, and A. Cichocki, "Optimization of SSVEP brain responses with application to eight-command brain–computer interface," *Neuroscience letters,* vol. 469, no. 1, pp. 34-38, 2010.

[20] J. R. Wolpaw, N. Birbaumer, D. J. McFarland, G. Pfurtscheller, and T. M. Vaughan, "Brain–computer interfaces for communication and control," *Clinical neurophysiology,* vol. 113, no. 6, pp. 767-791, 2002.

[21] C. Zhang *et al.*, "A survey on potentials, pathways and challenges of large language models in new-generation intelligent manufacturing," *Robotics and Computer-Integrated Manufacturing,* vol. 92, p. 102883, 2025.

[22] L. Shen, Y. Sun, Z. Yu, L. Ding, X. Tian, and D. Tao, "On Efficient Training of Large-Scale Deep Learning Models," *ACM Computing Surveys,* 2024.

[23] J. Achiam *et al.*, "Gpt-4 technical report," *arXiv preprint arXiv:2303.08774,* 2023.

[24] A. Chowdhery *et al.*, "Palm: Scaling language modeling with pathways," *Journal of Machine Learning Research,* vol. 24, no. 240, pp. 1-113, 2023.

[25] H. Liu, C. Li, Q. Wu, and Y. J. Lee, "Visual instruction tuning," *Advances in neural information processing systems,* vol. 36, 2024.

[26] B. Liu, X. Chen, N. Shi, Y. Wang, S. Gao, and X. Gao, "Improving the performance of individually calibrated SSVEP-BCI by task-discriminant component analysis," *IEEE Transactions on Neural Systems and Rehabilitation Engineering,* vol. 29, pp. 1998-2007, 2021.